\documentclass[aps,prl,superscriptaddress,twocolumn,longbibliography]{revtex4-1}
\usepackage{units}
\usepackage{amsmath}
\usepackage{amsthm}
\usepackage{amssymb}
\usepackage{graphicx}
\usepackage{color}
\usepackage[colorlinks=true, urlcolor=blue, citecolor=blue,linkcolor=blue,citebordercolor={1 0 0},linkbordercolor={0 0 1}]{hyperref}

\theoremstyle{plain}
\newtheorem{thm}{\protect\theoremname}
\theoremstyle{plain}

\ifx\proof\undefined
\newenvironment{proof}[1][\protect\proofname]{\par
	\normalfont\topsep6\p@\@plus6\p@\relax
	\trivlist
	\itemindent\parindent
	\item[\hskip\labelsep\scshape #1]\ignorespaces
}{%
	\endtrivlist\@endpefalse
}
\providecommand{\proofname}{Proof}
\fi

\makeatother

\providecommand{\factname}{Fact}
\providecommand{\theoremname}{Theorem}

\begin{document}
\title{Quantum Assisted Eigensolver}
\author{Kishor Bharti}
\email{kishor.bharti1@gmail.com}
\affiliation{Centre for Quantum Technologies, National University of Singapore 117543, Singapore}
\begin{abstract}
We propose a hybrid quantum-classical algorithm for approximating
the ground state and ground state energy of a Hamiltonian.
Once the Ansatz has been decided, the quantum part of the algorithm
involves the calculation of two overlap matrices. The output from the
quantum part of the algorithm is utilized as input for the classical computer. The classical part of the algorithm is a quadratically constrained
quadratic program with a single quadratic equality constraint. Unlike
the variational quantum eigensolver algorithm, our algorithm does not
have any classical-quantum feedback loop. Using convex relaxation
techniques, we provide an efficiently computable lower bound to the
classical optimization program. Furthermore, using results from Bar-On\textit{
et al. (Journal of Optimization Theory and Applications, 82(2):379--386,
1994}), we provide a sufficient condition for a local minimum to be
a global minimum. A solver can use such a condition as a stopping
criterion.
\end{abstract}
\maketitle

\noindent {\em Introduction.---}Most of the available quantum computers have at most $50-70$ noisy
qubits and nobody knows whether it is possible to utilise these noisy
intermediate-scale quantum (NISQ) \cite{preskill2018quantum} devices to solve any problem of
practical relevance. A recent experiment by researchers at Google
has, however, indicated the ability of these devices to execute computation
beyond the capability of any available powerful classical computer \cite{arute2019quantum}.
An important question is how such breakthroughs can be translated into quantum advantages for practical use-cases.

To harness the potential of NISQ era quantum computers, hybrid quantum-classical
variational algorithms have been suggested in recent years \cite{peruzzo2014variational,mcclean2016theory,kandala2017hardware,farhi2014quantum,farhi2016quantum}. These algorithms
operate via adaptive feedback control of the quantum device with the
objective function encoding the solution of the corresponding computational
problem. These algorithms are referred to as variational quantum algorithms (VQA).  Some of the canonical examples of VQAs are
variational quantum eigensolver (VQE) \cite{peruzzo2014variational,mcclean2016theory,kandala2017hardware}
and quantum approximate optimization algorithm (QAOA) \cite{farhi2014quantum,farhi2016quantum}.
The aforementioned algorithms employ classical optimizers to update the parameters in a parametric quantum circuit while utilizing the quantum computer to evaluate the objective function or its gradient efficiently. These algorithms are considered the best hope for a near-term quantum advantage.  The training of the parameters
of the parametric quantum circuit for these algorithms, in general, can be exceptionally challenging. The appearance of the vanishing
gradient problem (also referred to as barren plateau problem) as the
quantum hardware noise or circuit depth increases, has led to valid
concerns regarding the future of these algorithms \cite{mcclean2018barren,huang2019near,sharma2020trainability,cerezo2020cost,wang2020noise}. Moreover, the training landscape, in general,  does not correspond to any well-charactrized optimization program, thus making their investigation difficult.
Due to the aforementioned reasons, the pursuit of alternatives  becomes pertinent.

In this Letter, we propose a hybrid classical-quantum algorithm for
estimating the ground state of Hamiltonians. Unlike VQE,
our algorithm does not mandate tuning of parameters of a quantum circuit.
Moreover, there is no classical-quantum feedback loop. Our algorithm
proceeds in three stages, which we discuss informally now. The first
step involves the selection of a sufficiently expressible Ansatz. The
second step corresponds to the evaluation of specific matrix elements on
a quantum computer. We assume that the quantum computer can estimate the aforementioned matrix elements efficiently. After the second step, the
job of the quantum computer is over. The result from the second step
serves as input to the final step. The third step of our algorithm
is a classical optimization, which is a quadratically constrained
quadratic program (QCQP) with single quadratic equality constraint.

Using convex relaxation techniques such as Lagrangian relaxation and
semidefinite relaxation, we provide efficiently computable lower bounds
to the classical optimization program. We also provide a sufficient
condition for a local minimum to be a global minimum, which can be
used by a solver as a stopping criterion. We refer to our algorithm
as quantum assisted eigensolver (QAE). There has been a comprehensive study
of QCQPs in the non-convex optimization literature \cite{Boyd2004,bar1994global,d2003relaxations}
and one can hope to utilize many of the existing results from the aforementioned
community to prove theoretical guarantees for our algorithm. Our algorithm has connections with Ritz method, which is used to find an approximate solution for boundary value problems. We believe that studying the connections in detail will lead to improvements to our algorithm. Recently, many algorithms based on overlap matrix computation
have been suggested in the literature to estimate the ground state of Hamiltonians \cite{mcclean2017hybrid,kyriienko2020quantum,
parrish2019quantum,bespalova2020hamiltonian,huggins2020non,
takeshita2020increasing,stair2020multireference,
motta2020determining,seki2020quantum}. The second step of our algorithm could be improved based on ideas from the aforementioned references.

\medskip
{\noindent {\em Set-up---}}We assume that the Hamiltonian $H$ is a linear combination of unitaries, 
\begin{equation}
H=\sum_{i=1}^{n}\beta_{i}U_{i},\label{eq:ham_sum_unitaries}
\end{equation}
where $U_{i}\in SU\left(2^{N}\right)$, $n \in \mathbb{N}$, and $\beta_{i}\in\mathbb{C}$ for $i\in\{1,2,\cdots,n\}.$
Our goal is to calculate an approximation
to the ground state and ground state energy for $H.$ Let us consider
a set of $m$ quantum states $\left\{ \vert\phi_{j}\rangle\right\} _{j=1}^{m}$
such that $\langle\phi_{j}\vert\phi_{k}\rangle\neq1$ for $j\neq k$
and $\vert\phi_{j}\rangle=V_{j}\vert0^{\otimes^{N}}\rangle$, for
some efficiently implementable  unitary $V_{j}\in SU\left(2^{N}\right)$. Our Ansatz for approximating
the ground state is 
\begin{equation}
\vert\psi\left(\boldsymbol{\alpha}\right)\rangle=\sum_{i=1}^{m}\alpha_{j}\vert\phi_{j}\rangle,\label{eq:Ansatz}
\end{equation}
for $\boldsymbol{\alpha}\in\mathbb{C}^{m}.$ The task remains to find a
suitable $\hat{\boldsymbol{\alpha}}\in\mathbb{C}^{m}$ which minimises the expectation
value of $H$ for the Ansatz in \ref{eq:Ansatz}.

\medskip
{\noindent {\em Formulating the Optimization Problem---}}The expectation value of the Hamiltonian in \ref{eq:ham_sum_unitaries}
for the Ansatz in \ref{eq:Ansatz} can be expressed as
\begin{equation}
\left\langle H\left(\alpha\right)\right\rangle =\sum_{i,j,k}\beta_{i}\alpha_{j}\langle\phi_{j}\vert U_{i}\vert\phi_{k}\rangle\alpha_{k}.\label{eq:Expansion_Exp_1}
\end{equation}
Using the notation $D_{jk}\equiv\sum_{i}\beta_{i}\langle\phi_{j}\vert U_{i}\vert\phi_{k}\rangle$,
we get
\[
\left\langle H\left(\alpha\right)\right\rangle =\sum_{j,k}\alpha_{j}D_{j,k}\alpha_{k}
\]
\begin{equation}
=\boldsymbol{\alpha}^{T}D\boldsymbol{\alpha}.\label{eq:Expansion_Expect_2}
\end{equation}
The requirement on the Ansatz in \ref{eq:Ansatz} to be a valid quantum
state implies
\begin{equation}
\sum_{j,k}\alpha_{j}\langle\phi_{j}\vert\phi_{k}\rangle\alpha_{k}=1.\label{eq:Cons_1}
\end{equation}
Using the notation $E_{j,k}\equiv\langle\phi_{j}\vert\phi_{k}\rangle,$
we get
\begin{equation}
\sum_{j,k}\alpha_{j}E_{j,k}\alpha_{k}=1.\label{eq:cons_matrix}
\end{equation}
In compact notation, the above contraint can be re-written as
\begin{equation}
\boldsymbol{\alpha^{T}}E\boldsymbol{\alpha}=1.\label{eq:cons_compact}
\end{equation}
Combining \ref{eq:Expansion_Expect_2} and \ref{eq:cons_compact}, we get the following final optimization program, which we denote
by $P1.$
\[
\text{minimize }\boldsymbol{\alpha}^{T}D\boldsymbol{\alpha}
\]
\begin{equation}
\text{ subject to }\boldsymbol{\alpha^{T}}E\boldsymbol{\alpha}=1\label{eq:P1}
\end{equation}
The above program $P1$ is a quadratic optimization problem with a
single quadratic equality constraint. Furthermore, it is straightforward to see
that matrix $E$ is always a positive semidefinite matrix. The matrix
$D$ is positive semidefinite (PSD) whenever the Hamiltonian $H$
in \ref{eq:ham_sum_unitaries} is guaranteed to be PSD.

\medskip
{\noindent {\em The Hybrid Quantum-Classical Algorithm---}}For a given Hamiltonian (of form \ref{eq:ham_sum_unitaries}) the
QAE algorithm involves three steps:
\begin{enumerate}
\item Selection of an Ansatz,
\item Calculation of the overlap matrices ($D$ and $E$) on a quantum computer, and
\item Execution of the corresponding QCQP on a classical computer.
\end{enumerate}
Note that step $1$ is crucial and heavily determines the success
of the QAE algorithm. Our algorithm  demands the Ansatz to be a linear
combination of quantum states $\left\{ \vert\phi_{j}\rangle\right\} _{j=1}^{m}$
such that $\langle\phi_{j}\vert\phi_{k}\rangle\neq1$ for $j\neq k$
and $\vert\phi_{j}\rangle=V_{j}\vert0^{\otimes^{N}}\rangle$ for some
efficiently implementable unitary $V_{j}$.

Step $2$ involves the computation of the overlap matrices (D and E).  The computation of matrix elements of the form $\langle0^{\otimes N}\vert U\vert0^{\otimes N}\rangle$
(where $U$ is some unitary matrix) will furnish the overlap matrices  We assume that the quantum computer can efiiciently compute the aforementioned matrices for the given Hamiltonian and the choice of Ansatz.

\begin{figure}
\includegraphics[scale=0.25]{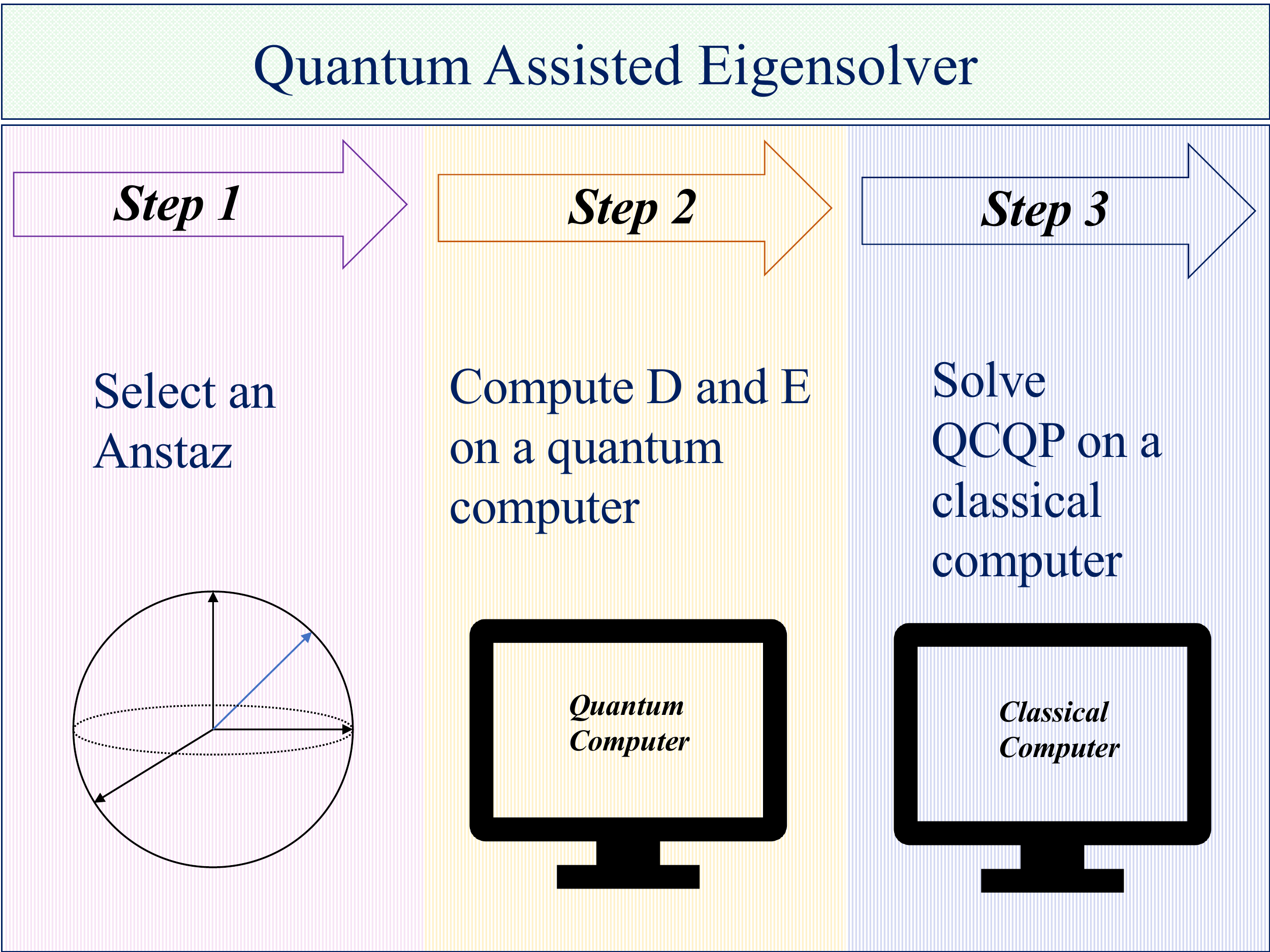}\caption{The QAE algorithm involves three steps. The first step corresponds
to selecting an efficiently preparable Ansatz. The second step concerns the
calculation of matrices $D$ and $E,$ where the elements of the matrices
are of the form $\langle0^{\otimes N}\vert U\vert0^{\otimes N}\rangle$
for some unitary $U\in SU\left(2^{N}\right).$ After step $2,$ we do not require
the quantum computer. The third step is an optimization program (more
precisely a QCQP), which is executed on a classical computer.}
\label{fig: QAE}
\end{figure}

Step $3$ corresponds to a QCQP with input matrices of size $m \times m.$
Note that the matrices $D$ and $E$ can be, in general, complex Hermitian.
The combination coefficients $\boldsymbol{\alpha}$ also in general
belong to $\mathbb{C}^{m}.$ However, one can always map the program
P1 with complex variables to a program with real variables, by creating
a mapping from $\mathbb{C}^{m}$ to $\mathbb{R}^{2m}$. We discuss
one such mapping in the Appendix. Thus, for the discussion
in the main text, without loss of generality, we will assume D and
E to be real symmetric matrices, and $\boldsymbol{\alpha}\in\mathbb{R}^{m}.$
We discuss the details of step $3$ in the follow-up paragraphs. See Figure \ref{fig: QAE} for a visual synopsis of the QAE algorithm.
If a given choice of Ansatz does not produce the desired accuracy,
one needs to revisit step $1$.

\medskip
{\noindent {\em Analysing the Classical Optimization---}}The objective function in $P1$ is convex whenever $D$ is PSD. The
quadratic equality in the constraint, however, renders the optimization
program $P1$ non-convex. Using the Lagrange multiplier $\lambda$
for the constraint $\boldsymbol{\alpha^{T}}E\boldsymbol{\alpha}=1,$
we define the following Lagrangian $\mathcal{L}:\mathbb{R}^{m}\times\mathbb{R}\rightarrow\mathbb{R},$
\begin{equation}
\mathcal{L}\left(\boldsymbol{\alpha},\lambda\right)=\boldsymbol{\alpha}^{T}D\boldsymbol{\alpha}+\lambda\left(\boldsymbol{\alpha^{T}}E\boldsymbol{\alpha}-1\right).\label{eq:Lag_01}
\end{equation}
The first order condition for the optimality of $P1$ are
\begin{equation}
\frac{\partial\mathcal{L}}{\partial\lambda}=0\implies\boldsymbol{\alpha^{T}}E\boldsymbol{\alpha}=1,\label{eq:First_order_1}
\end{equation}
and
\begin{equation}
\frac{\partial\mathcal{L}}{\partial\boldsymbol{\alpha}}=0\implies\left(D+\lambda E\right)\alpha=0.\label{eq:First_order_2}
\end{equation}
Using conditions \ref{eq:First_order_1} and \ref{eq:First_order_2},
we get
\textbf{
\begin{equation}
\boldsymbol{\alpha}^{T}D\boldsymbol{\alpha}+\lambda=0.\label{eq:first_order_final}
\end{equation}
}Using the tangent plane $T=\left\{ y:y^{T}E\boldsymbol{\alpha}=0\right\} $
and $H=D+\lambda E,$ the second order conditions for
optimality are given by 
\begin{equation}
y^{T}Hy\geq0\text{ }\forall y\in T.\label{eq:Second_order}
\end{equation}
\begin{thm}
Given a tuple $\left(\boldsymbol{\alpha},\lambda\right)$ that is
local minimum and satisfies $H\ge0$ yields the global minimum of
the program P1 in \ref{eq:P1}.
\end{thm}
\begin{proof}
The proof trivially follows from proposition 1 in \citep{bar1994global}
and has been deferred to Appendix.
\end{proof}
The aforementioned theorem provides a sufficient condition for a given
local minimum to be global minimum. Such a condition can be used by
a solver for P1 to provide a stopping criterion.

\medskip
{\noindent {\em Convex Relaxations---}}Lower bounds on the optimal value of the nonconvex optimization program
P1 can be provided by convex relaxation techniques for quadratically
constrained quadratic programs (QCQP). Minimizing over $\alpha$ for
the Lagrangian in \ref{eq:Lag_01}, we get the follwing dual function,
\[
g\left(\lambda\right)=\text{inf }\mathcal{L}\left(\alpha,\lambda\right)
\]
\begin{equation}
=- \lambda\text{ if }\left(D+\lambda E\right)\succcurlyeq0\label{eq:dual_function_1}
\end{equation}
and is unbounded otherwise. Using the dual function $g\left(\lambda\right),$
we can construct the following dual of P1:
\[
\text{maximize } \lambda
\]
\begin{equation}
\text{subject to }D-\lambda E\succcurlyeq0,\label{eq:Dual_P1}
\end{equation}
We will denote the above dual program in \ref{eq:Dual_P1} as D1.
Note that D1 is a semidefinite program (SDP) and hence can be solved
efficiently on a classical computer \citep{Boyd2004}. Instead of
Lagrangian relaxation (which turns out to be an SDP), one can also
have a direct SDP relaxation (see Appendix). The aforementioned convex
relaxation methods though provide a lower bound efficiently, the
optimal points in the convex relaxed program may not be feasible.
However, there exists sampling and convex-restriction based techniques
in the optimization theory literature, which provide good feasible
points \citep{d2003relaxations}.

\medskip
{\noindent {\em A Toy Example---}}We explain the implementation of our QAE algorithm by providing a
toy example. We consider the task of finding the ground state of the
Hydrogen molecule $\left(H_{2}\right)$. The Hamiltonian for $H_{2}$
can be represnted using two qubits as
\begin{equation}
H=a\left(Z\otimes I+I\otimes Z\right)+b\left(X\otimes X\right),\label{eq:Hyderogen_molecule}
\end{equation}
where $a=0.4$ and $b=0.2$ \citep{bravyi2017tapering,yamamoto2019natural}.
The symbols $X$ and $Z$ denote Pauli X and Pauli Z matrices. Our
first step is to select a set of efficiently preparable quantum states
to create the Ansatz (see \ref{eq:Ansatz}). This step is crucial
and will determine the expressibility of the Ansatz. For the sake of
explanation, we choose $\vert1,1\rangle,$$\vert+,-\rangle$ and $\vert-,+\rangle.$
For real coefficients $\alpha_{i}$ for $i\in\left\{ 1,2,3\right\} ,$
the Ansatz is given by
\begin{equation}
\vert\psi\left(\boldsymbol{\alpha}\right)\rangle=\alpha_{1}\vert1,1\rangle+\alpha_{2}\vert+,-\rangle+\alpha_{3}\vert-,+\rangle.\label{eq:Hydrogen_Ansatz}
\end{equation}
The expectation value of the Hydrogen molecule Hamiltonian in \ref{eq:Hyderogen_molecule}
for the Ansatz in \ref{eq:Hydrogen_Ansatz} can be written as $\boldsymbol{\alpha}^{T}D\boldsymbol{\alpha}$
(see \ref{eq:Expansion_Expect_2}). The normalization constraint for
the Ansatz can be written as $\boldsymbol{\alpha^{T}}E\boldsymbol{\alpha}=1$(see
\ref{eq:cons_compact}). For our toy example (with $a=0.4$
and $b=0.2$), we have
\begin{equation}
D=\left[\begin{array}{ccc}
-0.8 & 0.5 & 0.5\\
0.5 & -0.2 & 0\\
0.5 & 0 & -0.2
\end{array}\right]\label{eq:D_Hydrogen}
\end{equation}
and
\begin{equation}
E=\left[\begin{array}{ccc}
1 & -0.5 & -0.5\\
-0.5 & 1 & 0\\
-0.5 & 0 & 1
\end{array}\right].\label{eq:E_Hydrogen}
\end{equation}
Now the task remains to solve the optimization program, which has
to be done on a classical computer. Since, in our case, the Ansatz
coefficients $\left(\boldsymbol{\alpha}\in\mathbb{R}^{3}\right)$
and the input matrices ($D$ and $E$) are already real, no realification
is required. Solving the QCQP gives $\boldsymbol{\alpha}=[-0.87033111,0.1221769,0.1221776]^{T}.$
The corresponding ground state energy is $-0.824621,$ which is close
to the true ground state energy up to six digits after decimal$.$
Note that the success of the algorithm heavily depends on the Ansatz
and a bad Ansatz can give a poor result.

\medskip
{\noindent {\em Discussion and Open Problems---}In this work, we proposed a hybrid quantum-classical algorithm to
approximate the ground state and ground state energy of a Hamiltonian
of form \ref{eq:ham_sum_unitaries} via an Ansatz of form \ref{eq:Ansatz}.
Our algorithm proceeds in three steps. The first step involves selecting a set of efficiently preparable quantum states. The choice
of Ansatz heavily decides the proximity of the final output to
the correct value. Our algorithm requires the Ansatz to be a linear
combination of efficiently preparable distinct quantum states. In future, it would be interesting to develop
systematic as well as heuristic algorithms to provide sufficiently
expressible Ansatz which are compatible with our algorithm. The second
step involves the computation of matrix elements of the form $\langle0^{\otimes N}\vert U\vert0^{\otimes N}\rangle$
for some unitary $U$. Note that the aforementioned matrix elements can
have complex values and a straightforward execution of the unitary
$U$ on a circuit initialized in $\vert0^{\otimes N}\rangle$, followed
by measurement in the computational basis will kill all the phase
information. One popolar approach to calculate $\langle0^{\otimes N}\vert U\vert0^{\otimes N}\rangle$
is the Hadamard test, which involves controlled-unitaries. Since implementing controlled-unitaries is challenging, one needs to devise alternative methods. Ideas from  \cite{mitarai2019methodology} could be a good starting point to proceed in the aforementioned direction. Once the matrix elements have been computed, the
role of the quantum computer in the algorithm is over. Unlike VQEs,
there is no feedback loop between the classical and quantum computer.
The output from the first step is used to provide input matrices ($D$
and $E$) for the second step of the algorithm. The third step corresponds
to a non-convex optimization program (QCQP) with input matrices of
size $m \times m .$ Using standard techniques in optimization
theory, we provided a convex relaxation to the problem, which can
be used to efficiently compute a lower bound. Note that the aforementioned
lower bound corresponds to the lower bound for the given choice of
Ansatz. The optimal points in the convex relaxed program may not be
a feasible point. However, one can use sampling and convex-restriction
based techniques in the optimization theory literature to find good
enough feasible points \cite{d2003relaxations}. Furthermore, we
discussed the sufficient condition for a local minimum to be a global
minimum. There has been a thorough study of QCQPs in the non-convex optimization
literature, and one can hope to use many of the existing results for QCPQs
to prove theoretical guarantees for our
algorithm. 
The noise in the quantum system can affect the computation of $D$
and $E$ matrices on a quantum computer. In future, it would be interesting
to study the effect of noise on our algorithm. To comprehend the potential
of our algorithm, it would be interesting to deliver theoretical guarantees
using tools from complexity theory. In recent years, a series of results
have been supplied in the VQE and QAOA set-up where the key idea has
been to map the problem to finding the ground state of a Hamiltonian \cite{farhi2014quantum,anschuetz2019variational,garcia2018addressing}. It would be
interesting to study these problems in our set-up. We hope that such
studies can lead to the demonstration of quantum advantage for useful
problems on NISQ devices.

\medskip
{\noindent {\em Acknowledgements---}}We thank Patrick Rebentrost, Atul Singh Arora, Antonios Varvitsiotis
and Tobias Haug for interesting discussions. We are grateful to the
National Research Foundation and the Ministry of Education, Singapore
for financial support. We thank Mohua Das for support with the pictorial description of the QAE algorithm.

\bibliographystyle{apsrev4-1}
\bibliography{QAE}

\begin{thebibliography}{29}%
\makeatletter
\providecommand \@ifxundefined [1]{%
 \@ifx{#1\undefined}
}%
\providecommand \@ifnum [1]{%
 \ifnum #1\expandafter \@firstoftwo
 \else \expandafter \@secondoftwo
 \fi
}%
\providecommand \@ifx [1]{%
 \ifx #1\expandafter \@firstoftwo
 \else \expandafter \@secondoftwo
 \fi
}%
\providecommand \natexlab [1]{#1}%
\providecommand \enquote  [1]{``#1''}%
\providecommand \bibnamefont  [1]{#1}%
\providecommand \bibfnamefont [1]{#1}%
\providecommand \citenamefont [1]{#1}%
\providecommand \href@noop [0]{\@secondoftwo}%
\providecommand \href [0]{\begingroup \@sanitize@url \@href}%
\providecommand \@href[1]{\@@startlink{#1}\@@href}%
\providecommand \@@href[1]{\endgroup#1\@@endlink}%
\providecommand \@sanitize@url [0]{\catcode `\\12\catcode `\$12\catcode
  `\&12\catcode `\#12\catcode `\^12\catcode `\_12\catcode `\%12\relax}%
\providecommand \@@startlink[1]{}%
\providecommand \@@endlink[0]{}%
\providecommand \url  [0]{\begingroup\@sanitize@url \@url }%
\providecommand \@url [1]{\endgroup\@href {#1}{\urlprefix }}%
\providecommand \urlprefix  [0]{URL }%
\providecommand \Eprint [0]{\href }%
\providecommand \doibase [0]{http://dx.doi.org/}%
\providecommand \selectlanguage [0]{\@gobble}%
\providecommand \bibinfo  [0]{\@secondoftwo}%
\providecommand \bibfield  [0]{\@secondoftwo}%
\providecommand \translation [1]{[#1]}%
\providecommand \BibitemOpen [0]{}%
\providecommand \bibitemStop [0]{}%
\providecommand \bibitemNoStop [0]{.\EOS\space}%
\providecommand \EOS [0]{\spacefactor3000\relax}%
\providecommand \BibitemShut  [1]{\csname bibitem#1\endcsname}%
\let\auto@bib@innerbib\@empty
%</preamble>
\bibitem [{\citenamefont {Preskill}(2018)}]{preskill2018quantum}%
  \BibitemOpen
  \bibfield  {author} {\bibinfo {author} {\bibfnamefont {J.}~\bibnamefont
  {Preskill}},\ }\href@noop {} {\bibfield  {journal} {\bibinfo  {journal}
  {Quantum}\ }\textbf {\bibinfo {volume} {2}},\ \bibinfo {pages} {79} (\bibinfo
  {year} {2018})}\BibitemShut {NoStop}%
\bibitem [{\citenamefont {Arute}\ \emph {et~al.}(2019)\citenamefont {Arute},
  \citenamefont {Arya}, \citenamefont {Babbush}, \citenamefont {Bacon},
  \citenamefont {Bardin}, \citenamefont {Barends}, \citenamefont {Biswas},
  \citenamefont {Boixo}, \citenamefont {Brandao}, \citenamefont {Buell} \emph
  {et~al.}}]{arute2019quantum}%
  \BibitemOpen
  \bibfield  {author} {\bibinfo {author} {\bibfnamefont {F.}~\bibnamefont
  {Arute}}, \bibinfo {author} {\bibfnamefont {K.}~\bibnamefont {Arya}},
  \bibinfo {author} {\bibfnamefont {R.}~\bibnamefont {Babbush}}, \bibinfo
  {author} {\bibfnamefont {D.}~\bibnamefont {Bacon}}, \bibinfo {author}
  {\bibfnamefont {J.~C.}\ \bibnamefont {Bardin}}, \bibinfo {author}
  {\bibfnamefont {R.}~\bibnamefont {Barends}}, \bibinfo {author} {\bibfnamefont
  {R.}~\bibnamefont {Biswas}}, \bibinfo {author} {\bibfnamefont
  {S.}~\bibnamefont {Boixo}}, \bibinfo {author} {\bibfnamefont {F.~G.}\
  \bibnamefont {Brandao}}, \bibinfo {author} {\bibfnamefont {D.~A.}\
  \bibnamefont {Buell}},  \emph {et~al.},\ }\href@noop {} {\bibfield  {journal}
  {\bibinfo  {journal} {Nature}\ }\textbf {\bibinfo {volume} {574}},\ \bibinfo
  {pages} {505} (\bibinfo {year} {2019})}\BibitemShut {NoStop}%
\bibitem [{\citenamefont {Peruzzo}\ \emph {et~al.}(2014)\citenamefont
  {Peruzzo}, \citenamefont {McClean}, \citenamefont {Shadbolt}, \citenamefont
  {Yung}, \citenamefont {Zhou}, \citenamefont {Love}, \citenamefont
  {Aspuru-Guzik},\ and\ \citenamefont {Obrien}}]{peruzzo2014variational}%
  \BibitemOpen
  \bibfield  {author} {\bibinfo {author} {\bibfnamefont {A.}~\bibnamefont
  {Peruzzo}}, \bibinfo {author} {\bibfnamefont {J.}~\bibnamefont {McClean}},
  \bibinfo {author} {\bibfnamefont {P.}~\bibnamefont {Shadbolt}}, \bibinfo
  {author} {\bibfnamefont {M.-H.}\ \bibnamefont {Yung}}, \bibinfo {author}
  {\bibfnamefont {X.-Q.}\ \bibnamefont {Zhou}}, \bibinfo {author}
  {\bibfnamefont {P.~J.}\ \bibnamefont {Love}}, \bibinfo {author}
  {\bibfnamefont {A.}~\bibnamefont {Aspuru-Guzik}}, \ and\ \bibinfo {author}
  {\bibfnamefont {J.~L.}\ \bibnamefont {Obrien}},\ }\href@noop {} {\bibfield
  {journal} {\bibinfo  {journal} {Nature communications}\ }\textbf {\bibinfo
  {volume} {5}},\ \bibinfo {pages} {4213} (\bibinfo {year} {2014})}\BibitemShut
  {NoStop}%
\bibitem [{\citenamefont {McClean}\ \emph {et~al.}(2016)\citenamefont
  {McClean}, \citenamefont {Romero}, \citenamefont {Babbush},\ and\
  \citenamefont {Aspuru-Guzik}}]{mcclean2016theory}%
  \BibitemOpen
  \bibfield  {author} {\bibinfo {author} {\bibfnamefont {J.~R.}\ \bibnamefont
  {McClean}}, \bibinfo {author} {\bibfnamefont {J.}~\bibnamefont {Romero}},
  \bibinfo {author} {\bibfnamefont {R.}~\bibnamefont {Babbush}}, \ and\
  \bibinfo {author} {\bibfnamefont {A.}~\bibnamefont {Aspuru-Guzik}},\
  }\href@noop {} {\bibfield  {journal} {\bibinfo  {journal} {New Journal of
  Physics}\ }\textbf {\bibinfo {volume} {18}},\ \bibinfo {pages} {023023}
  (\bibinfo {year} {2016})}\BibitemShut {NoStop}%
\bibitem [{\citenamefont {Kandala}\ \emph {et~al.}(2017)\citenamefont
  {Kandala}, \citenamefont {Mezzacapo}, \citenamefont {Temme}, \citenamefont
  {Takita}, \citenamefont {Brink}, \citenamefont {Chow},\ and\ \citenamefont
  {Gambetta}}]{kandala2017hardware}%
  \BibitemOpen
  \bibfield  {author} {\bibinfo {author} {\bibfnamefont {A.}~\bibnamefont
  {Kandala}}, \bibinfo {author} {\bibfnamefont {A.}~\bibnamefont {Mezzacapo}},
  \bibinfo {author} {\bibfnamefont {K.}~\bibnamefont {Temme}}, \bibinfo
  {author} {\bibfnamefont {M.}~\bibnamefont {Takita}}, \bibinfo {author}
  {\bibfnamefont {M.}~\bibnamefont {Brink}}, \bibinfo {author} {\bibfnamefont
  {J.~M.}\ \bibnamefont {Chow}}, \ and\ \bibinfo {author} {\bibfnamefont
  {J.~M.}\ \bibnamefont {Gambetta}},\ }\href@noop {} {\bibfield  {journal}
  {\bibinfo  {journal} {Nature}\ }\textbf {\bibinfo {volume} {549}},\ \bibinfo
  {pages} {242} (\bibinfo {year} {2017})}\BibitemShut {NoStop}%
\bibitem [{\citenamefont {Farhi}\ \emph {et~al.}(2014)\citenamefont {Farhi},
  \citenamefont {Goldstone},\ and\ \citenamefont {Gutmann}}]{farhi2014quantum}%
  \BibitemOpen
  \bibfield  {author} {\bibinfo {author} {\bibfnamefont {E.}~\bibnamefont
  {Farhi}}, \bibinfo {author} {\bibfnamefont {J.}~\bibnamefont {Goldstone}}, \
  and\ \bibinfo {author} {\bibfnamefont {S.}~\bibnamefont {Gutmann}},\
  }\href@noop {} {\bibfield  {journal} {\bibinfo  {journal} {arXiv preprint
  arXiv:1411.4028}\ } (\bibinfo {year} {2014})}\BibitemShut {NoStop}%
\bibitem [{\citenamefont {Farhi}\ and\ \citenamefont
  {Harrow}(2016)}]{farhi2016quantum}%
  \BibitemOpen
  \bibfield  {author} {\bibinfo {author} {\bibfnamefont {E.}~\bibnamefont
  {Farhi}}\ and\ \bibinfo {author} {\bibfnamefont {A.~W.}\ \bibnamefont
  {Harrow}},\ }\href@noop {} {\bibfield  {journal} {\bibinfo  {journal} {arXiv
  preprint arXiv:1602.07674}\ } (\bibinfo {year} {2016})}\BibitemShut {NoStop}%
\bibitem [{\citenamefont {McClean}\ \emph {et~al.}(2018)\citenamefont
  {McClean}, \citenamefont {Boixo}, \citenamefont {Smelyanskiy}, \citenamefont
  {Babbush},\ and\ \citenamefont {Neven}}]{mcclean2018barren}%
  \BibitemOpen
  \bibfield  {author} {\bibinfo {author} {\bibfnamefont {J.~R.}\ \bibnamefont
  {McClean}}, \bibinfo {author} {\bibfnamefont {S.}~\bibnamefont {Boixo}},
  \bibinfo {author} {\bibfnamefont {V.~N.}\ \bibnamefont {Smelyanskiy}},
  \bibinfo {author} {\bibfnamefont {R.}~\bibnamefont {Babbush}}, \ and\
  \bibinfo {author} {\bibfnamefont {H.}~\bibnamefont {Neven}},\ }\href@noop {}
  {\bibfield  {journal} {\bibinfo  {journal} {Nature communications}\ }\textbf
  {\bibinfo {volume} {9}},\ \bibinfo {pages} {4812} (\bibinfo {year}
  {2018})}\BibitemShut {NoStop}%
\bibitem [{\citenamefont {Huang}\ \emph {et~al.}(2019)\citenamefont {Huang},
  \citenamefont {Bharti},\ and\ \citenamefont {Rebentrost}}]{huang2019near}%
  \BibitemOpen
  \bibfield  {author} {\bibinfo {author} {\bibfnamefont {H.-Y.}\ \bibnamefont
  {Huang}}, \bibinfo {author} {\bibfnamefont {K.}~\bibnamefont {Bharti}}, \
  and\ \bibinfo {author} {\bibfnamefont {P.}~\bibnamefont {Rebentrost}},\
  }\href@noop {} {\bibfield  {journal} {\bibinfo  {journal} {arXiv preprint
  arXiv:1909.07344}\ } (\bibinfo {year} {2019})}\BibitemShut {NoStop}%
\bibitem [{\citenamefont {Sharma}\ \emph {et~al.}(2020)\citenamefont {Sharma},
  \citenamefont {Cerezo}, \citenamefont {Cincio},\ and\ \citenamefont
  {Coles}}]{sharma2020trainability}%
  \BibitemOpen
  \bibfield  {author} {\bibinfo {author} {\bibfnamefont {K.}~\bibnamefont
  {Sharma}}, \bibinfo {author} {\bibfnamefont {M.}~\bibnamefont {Cerezo}},
  \bibinfo {author} {\bibfnamefont {L.}~\bibnamefont {Cincio}}, \ and\ \bibinfo
  {author} {\bibfnamefont {P.~J.}\ \bibnamefont {Coles}},\ }\href@noop {}
  {\bibfield  {journal} {\bibinfo  {journal} {arXiv preprint arXiv:2005.12458}\
  } (\bibinfo {year} {2020})}\BibitemShut {NoStop}%
\bibitem [{\citenamefont {Cerezo}\ \emph {et~al.}(2020)\citenamefont {Cerezo},
  \citenamefont {Sone}, \citenamefont {Volkoff}, \citenamefont {Cincio},\ and\
  \citenamefont {Coles}}]{cerezo2020cost}%
  \BibitemOpen
  \bibfield  {author} {\bibinfo {author} {\bibfnamefont {M.}~\bibnamefont
  {Cerezo}}, \bibinfo {author} {\bibfnamefont {A.}~\bibnamefont {Sone}},
  \bibinfo {author} {\bibfnamefont {T.}~\bibnamefont {Volkoff}}, \bibinfo
  {author} {\bibfnamefont {L.}~\bibnamefont {Cincio}}, \ and\ \bibinfo {author}
  {\bibfnamefont {P.~J.}\ \bibnamefont {Coles}},\ }\href@noop {} {\bibfield
  {journal} {\bibinfo  {journal} {arXiv preprint arXiv:2001.00550}\ } (\bibinfo
  {year} {2020})}\BibitemShut {NoStop}%
\bibitem [{\citenamefont {Wang}\ \emph {et~al.}(2020)\citenamefont {Wang},
  \citenamefont {Fontana}, \citenamefont {Cerezo}, \citenamefont {Sharma},
  \citenamefont {Sone}, \citenamefont {Cincio},\ and\ \citenamefont
  {Coles}}]{wang2020noise}%
  \BibitemOpen
  \bibfield  {author} {\bibinfo {author} {\bibfnamefont {S.}~\bibnamefont
  {Wang}}, \bibinfo {author} {\bibfnamefont {E.}~\bibnamefont {Fontana}},
  \bibinfo {author} {\bibfnamefont {M.}~\bibnamefont {Cerezo}}, \bibinfo
  {author} {\bibfnamefont {K.}~\bibnamefont {Sharma}}, \bibinfo {author}
  {\bibfnamefont {A.}~\bibnamefont {Sone}}, \bibinfo {author} {\bibfnamefont
  {L.}~\bibnamefont {Cincio}}, \ and\ \bibinfo {author} {\bibfnamefont {P.~J.}\
  \bibnamefont {Coles}},\ }\href@noop {} {\bibfield  {journal} {\bibinfo
  {journal} {arXiv preprint arXiv:2007.14384}\ } (\bibinfo {year}
  {2020})}\BibitemShut {NoStop}%
\bibitem [{\citenamefont {Boyd}\ and\ \citenamefont
  {Vandenberghe}(2004)}]{Boyd2004}%
  \BibitemOpen
  \bibfield  {author} {\bibinfo {author} {\bibfnamefont {S.}~\bibnamefont
  {Boyd}}\ and\ \bibinfo {author} {\bibfnamefont {L.}~\bibnamefont
  {Vandenberghe}},\ }\href@noop {} {\emph {\bibinfo {title} {Convex
  Optimization}}}\ (\bibinfo  {publisher} {Cambridge University Press},\
  \bibinfo {year} {2004})\BibitemShut {NoStop}%
\bibitem [{\citenamefont {Bar-On}\ and\ \citenamefont
  {Grasse}(1994)}]{bar1994global}%
  \BibitemOpen
  \bibfield  {author} {\bibinfo {author} {\bibfnamefont {J.}~\bibnamefont
  {Bar-On}}\ and\ \bibinfo {author} {\bibfnamefont {K.}~\bibnamefont
  {Grasse}},\ }\href@noop {} {\bibfield  {journal} {\bibinfo  {journal}
  {Journal of Optimization Theory and Applications}\ }\textbf {\bibinfo
  {volume} {82}},\ \bibinfo {pages} {379} (\bibinfo {year} {1994})}\BibitemShut
  {NoStop}%
\bibitem [{\citenamefont {dAspremont}\ and\ \citenamefont
  {Boyd}(2003)}]{d2003relaxations}%
  \BibitemOpen
  \bibfield  {author} {\bibinfo {author} {\bibfnamefont {A.}~\bibnamefont
  {dAspremont}}\ and\ \bibinfo {author} {\bibfnamefont {S.}~\bibnamefont
  {Boyd}},\ }\href@noop {} {\bibfield  {journal} {\bibinfo  {journal} {EE392o
  Class Notes, Stanford University}\ }\textbf {\bibinfo {volume} {1}},\
  \bibinfo {pages} {1} (\bibinfo {year} {2003})}\BibitemShut {NoStop}%
\bibitem [{\citenamefont {McClean}\ \emph {et~al.}(2017)\citenamefont
  {McClean}, \citenamefont {Kimchi-Schwartz}, \citenamefont {Carter},\ and\
  \citenamefont {De~Jong}}]{mcclean2017hybrid}%
  \BibitemOpen
  \bibfield  {author} {\bibinfo {author} {\bibfnamefont {J.~R.}\ \bibnamefont
  {McClean}}, \bibinfo {author} {\bibfnamefont {M.~E.}\ \bibnamefont
  {Kimchi-Schwartz}}, \bibinfo {author} {\bibfnamefont {J.}~\bibnamefont
  {Carter}}, \ and\ \bibinfo {author} {\bibfnamefont {W.~A.}\ \bibnamefont
  {De~Jong}},\ }\href@noop {} {\bibfield  {journal} {\bibinfo  {journal}
  {Physical Review A}\ }\textbf {\bibinfo {volume} {95}},\ \bibinfo {pages}
  {042308} (\bibinfo {year} {2017})}\BibitemShut {NoStop}%
\bibitem [{\citenamefont {Kyriienko}(2020)}]{kyriienko2020quantum}%
  \BibitemOpen
  \bibfield  {author} {\bibinfo {author} {\bibfnamefont {O.}~\bibnamefont
  {Kyriienko}},\ }\href@noop {} {\bibfield  {journal} {\bibinfo  {journal} {npj
  Quantum Information}\ }\textbf {\bibinfo {volume} {6}},\ \bibinfo {pages} {1}
  (\bibinfo {year} {2020})}\BibitemShut {NoStop}%
\bibitem [{\citenamefont {Parrish}\ and\ \citenamefont
  {McMahon}(2019)}]{parrish2019quantum}%
  \BibitemOpen
  \bibfield  {author} {\bibinfo {author} {\bibfnamefont {R.~M.}\ \bibnamefont
  {Parrish}}\ and\ \bibinfo {author} {\bibfnamefont {P.~L.}\ \bibnamefont
  {McMahon}},\ }\href@noop {} {\bibfield  {journal} {\bibinfo  {journal} {arXiv
  preprint arXiv: 1909.08925}\ } (\bibinfo {year} {2019})}\BibitemShut
  {NoStop}%
\bibitem [{\citenamefont {Bespalova}\ and\ \citenamefont
  {Kyriienko}(2020)}]{bespalova2020hamiltonian}%
  \BibitemOpen
  \bibfield  {author} {\bibinfo {author} {\bibfnamefont {T.~A.}\ \bibnamefont
  {Bespalova}}\ and\ \bibinfo {author} {\bibfnamefont {O.}~\bibnamefont
  {Kyriienko}},\ }\href@noop {} {\bibfield  {journal} {\bibinfo  {journal}
  {arXiv preprint arXiv:2009.03351}\ } (\bibinfo {year} {2020})}\BibitemShut
  {NoStop}%
\bibitem [{\citenamefont {Huggins}\ \emph {et~al.}(2020)\citenamefont
  {Huggins}, \citenamefont {Lee}, \citenamefont {Baek}, \citenamefont
  {O'Gorman},\ and\ \citenamefont {Whaley}}]{huggins2020non}%
  \BibitemOpen
  \bibfield  {author} {\bibinfo {author} {\bibfnamefont {W.~J.}\ \bibnamefont
  {Huggins}}, \bibinfo {author} {\bibfnamefont {J.}~\bibnamefont {Lee}},
  \bibinfo {author} {\bibfnamefont {U.}~\bibnamefont {Baek}}, \bibinfo {author}
  {\bibfnamefont {B.}~\bibnamefont {O'Gorman}}, \ and\ \bibinfo {author}
  {\bibfnamefont {K.~B.}\ \bibnamefont {Whaley}},\ }\href@noop {} {\bibfield
  {journal} {\bibinfo  {journal} {New Journal of Physics}\ } (\bibinfo {year}
  {2020})}\BibitemShut {NoStop}%
\bibitem [{\citenamefont {Takeshita}\ \emph {et~al.}(2020)\citenamefont
  {Takeshita}, \citenamefont {Rubin}, \citenamefont {Jiang}, \citenamefont
  {Lee}, \citenamefont {Babbush},\ and\ \citenamefont
  {McClean}}]{takeshita2020increasing}%
  \BibitemOpen
  \bibfield  {author} {\bibinfo {author} {\bibfnamefont {T.}~\bibnamefont
  {Takeshita}}, \bibinfo {author} {\bibfnamefont {N.~C.}\ \bibnamefont
  {Rubin}}, \bibinfo {author} {\bibfnamefont {Z.}~\bibnamefont {Jiang}},
  \bibinfo {author} {\bibfnamefont {E.}~\bibnamefont {Lee}}, \bibinfo {author}
  {\bibfnamefont {R.}~\bibnamefont {Babbush}}, \ and\ \bibinfo {author}
  {\bibfnamefont {J.~R.}\ \bibnamefont {McClean}},\ }\href@noop {} {\bibfield
  {journal} {\bibinfo  {journal} {Physical Review X}\ }\textbf {\bibinfo
  {volume} {10}},\ \bibinfo {pages} {011004} (\bibinfo {year}
  {2020})}\BibitemShut {NoStop}%
\bibitem [{\citenamefont {Stair}\ \emph {et~al.}(2020)\citenamefont {Stair},
  \citenamefont {Huang},\ and\ \citenamefont
  {Evangelista}}]{stair2020multireference}%
  \BibitemOpen
  \bibfield  {author} {\bibinfo {author} {\bibfnamefont {N.~H.}\ \bibnamefont
  {Stair}}, \bibinfo {author} {\bibfnamefont {R.}~\bibnamefont {Huang}}, \ and\
  \bibinfo {author} {\bibfnamefont {F.~A.}\ \bibnamefont {Evangelista}},\
  }\href@noop {} {\bibfield  {journal} {\bibinfo  {journal} {Journal of
  Chemical Theory and Computation}\ }\textbf {\bibinfo {volume} {16}},\
  \bibinfo {pages} {2236} (\bibinfo {year} {2020})}\BibitemShut {NoStop}%
\bibitem [{\citenamefont {Motta}\ \emph {et~al.}(2020)\citenamefont {Motta},
  \citenamefont {Sun}, \citenamefont {Tan}, \citenamefont {O’Rourke},
  \citenamefont {Ye}, \citenamefont {Minnich}, \citenamefont {Brand{\~a}o},\
  and\ \citenamefont {Chan}}]{motta2020determining}%
  \BibitemOpen
  \bibfield  {author} {\bibinfo {author} {\bibfnamefont {M.}~\bibnamefont
  {Motta}}, \bibinfo {author} {\bibfnamefont {C.}~\bibnamefont {Sun}}, \bibinfo
  {author} {\bibfnamefont {A.~T.}\ \bibnamefont {Tan}}, \bibinfo {author}
  {\bibfnamefont {M.~J.}\ \bibnamefont {O’Rourke}}, \bibinfo {author}
  {\bibfnamefont {E.}~\bibnamefont {Ye}}, \bibinfo {author} {\bibfnamefont
  {A.~J.}\ \bibnamefont {Minnich}}, \bibinfo {author} {\bibfnamefont {F.~G.}\
  \bibnamefont {Brand{\~a}o}}, \ and\ \bibinfo {author} {\bibfnamefont
  {G.~K.-L.}\ \bibnamefont {Chan}},\ }\href@noop {} {\bibfield  {journal}
  {\bibinfo  {journal} {Nature Physics}\ }\textbf {\bibinfo {volume} {16}},\
  \bibinfo {pages} {205} (\bibinfo {year} {2020})}\BibitemShut {NoStop}%
\bibitem [{\citenamefont {Seki}\ and\ \citenamefont
  {Yunoki}(2020)}]{seki2020quantum}%
  \BibitemOpen
  \bibfield  {author} {\bibinfo {author} {\bibfnamefont {K.}~\bibnamefont
  {Seki}}\ and\ \bibinfo {author} {\bibfnamefont {S.}~\bibnamefont {Yunoki}},\
  }\href@noop {} {\bibfield  {journal} {\bibinfo  {journal} {arXiv:2008.03661}\
  } (\bibinfo {year} {2020})}\BibitemShut {NoStop}%
\bibitem [{\citenamefont {Bravyi}\ \emph {et~al.}(2017)\citenamefont {Bravyi},
  \citenamefont {Gambetta}, \citenamefont {Mezzacapo},\ and\ \citenamefont
  {Temme}}]{bravyi2017tapering}%
  \BibitemOpen
  \bibfield  {author} {\bibinfo {author} {\bibfnamefont {S.}~\bibnamefont
  {Bravyi}}, \bibinfo {author} {\bibfnamefont {J.~M.}\ \bibnamefont
  {Gambetta}}, \bibinfo {author} {\bibfnamefont {A.}~\bibnamefont {Mezzacapo}},
  \ and\ \bibinfo {author} {\bibfnamefont {K.}~\bibnamefont {Temme}},\
  }\href@noop {} {\bibfield  {journal} {\bibinfo  {journal} {arXiv preprint
  arXiv:1701.08213}\ } (\bibinfo {year} {2017})}\BibitemShut {NoStop}%
\bibitem [{\citenamefont {Yamamoto}(2019)}]{yamamoto2019natural}%
  \BibitemOpen
  \bibfield  {author} {\bibinfo {author} {\bibfnamefont {N.}~\bibnamefont
  {Yamamoto}},\ }\href@noop {} {\bibfield  {journal} {\bibinfo  {journal}
  {arXiv preprint arXiv:1909.05074}\ } (\bibinfo {year} {2019})}\BibitemShut
  {NoStop}%
\bibitem [{\citenamefont {Mitarai}\ and\ \citenamefont
  {Fujii}(2019)}]{mitarai2019methodology}%
  \BibitemOpen
  \bibfield  {author} {\bibinfo {author} {\bibfnamefont {K.}~\bibnamefont
  {Mitarai}}\ and\ \bibinfo {author} {\bibfnamefont {K.}~\bibnamefont
  {Fujii}},\ }\href@noop {} {\bibfield  {journal} {\bibinfo  {journal}
  {Physical Review Research}\ }\textbf {\bibinfo {volume} {1}},\ \bibinfo
  {pages} {013006} (\bibinfo {year} {2019})}\BibitemShut {NoStop}%
\bibitem [{\citenamefont {Anschuetz}\ \emph {et~al.}(2019)\citenamefont
  {Anschuetz}, \citenamefont {Olson}, \citenamefont {Aspuru-Guzik},\ and\
  \citenamefont {Cao}}]{anschuetz2019variational}%
  \BibitemOpen
  \bibfield  {author} {\bibinfo {author} {\bibfnamefont {E.}~\bibnamefont
  {Anschuetz}}, \bibinfo {author} {\bibfnamefont {J.}~\bibnamefont {Olson}},
  \bibinfo {author} {\bibfnamefont {A.}~\bibnamefont {Aspuru-Guzik}}, \ and\
  \bibinfo {author} {\bibfnamefont {Y.}~\bibnamefont {Cao}},\ }in\ \href@noop
  {} {\emph {\bibinfo {booktitle} {International Workshop on Quantum Technology
  and Optimization Problems}}}\ (\bibinfo {organization} {Springer},\ \bibinfo
  {year} {2019})\ pp.\ \bibinfo {pages} {74--85}\BibitemShut {NoStop}%
\bibitem [{\citenamefont {Garcia-Saez}\ and\ \citenamefont
  {Latorre}(2018)}]{garcia2018addressing}%
  \BibitemOpen
  \bibfield  {author} {\bibinfo {author} {\bibfnamefont {A.}~\bibnamefont
  {Garcia-Saez}}\ and\ \bibinfo {author} {\bibfnamefont {J.}~\bibnamefont
  {Latorre}},\ }\href@noop {} {\bibfield  {journal} {\bibinfo  {journal} {arXiv
  preprint arXiv:1806.02287}\ } (\bibinfo {year} {2018})}\BibitemShut {NoStop}%
\end{thebibliography}%

\appendix 
\section{Inequality Representation\label{sec:Inequality-Representation}}
The equality constraint in P1 can also be rewritten as two inequality
constraints. In terms of inequality constraints, the program P1 becomes:
\[
\text{minimize }\alpha^{T}D\alpha
\]
\begin{equation}
\text{subject to }\alpha^{T}E\alpha-1\leq0,\label{eq:P2}
\end{equation}
\[
\alpha^{T}(-E)\alpha+1\leq0.
\]
We denote the above QCQP by P2. 
\section{Semidefinite Relaxation\label{sec:Semidefinite-Relaxation}}

In the main text, we discussed the Lagrangian relaxation of the QCQP
in \ref{eq:P1} i.e., P1. However, one can also have a direct convex relaxation
using SDP. The SDP relaxation of program in \ref{eq:P1} can be written as 
\[
\text{minimize}\text{ Tr}\left(XD\right)
\]
\begin{equation}
\text{subject to }\text{Tr}\left(XE\right)=1,\label{eq:Direct_SDP_Relaxation}
\end{equation}
\[
\left[\begin{array}{cc}
X & x\\
x^{T} & 1
\end{array}\right]\succcurlyeq0,
\]
\[
X\in\mathbb{S}^{m\times m},x\in\mathbb{R}^{1\times m}.
\]
Here $X$ is a symmetric matrix of size $m\times m$ and $x$ is a
row vector of size $1\times m.$
\section{Realification\label{sec:Realification}}
The idea of realification is to create a mapping from $\mathbb{C}^{n}$
to $\mathbb{R}^{2n}.$ The crux of the idea is the following real
matrix representation for $1$ and $i$,
\begin{equation}
1\leftrightarrow\left[\begin{array}{cc}
1 & 0\\
0 & 1
\end{array}\right]\equiv\mathbb{I}\label{eq:map_1}
\end{equation}
and
\begin{equation}
i\leftrightarrow\left[\begin{array}{cc}
0 & -1\\
1 & 0
\end{array}\right]\equiv I.\label{eq:map_iota}
\end{equation}
It can be seen that $I^{2}=-\mathbb{I},$which mimics $i^{2}=-1.$
Using the mappings in \ref{eq:map_1} and \ref{eq:map_iota}, one
can obtain the following representation with a double-sided implication
on positivity,
\begin{equation}
M\succcurlyeq0\iff\left[\begin{array}{cc}
M_{R} & -M_{I}\\
M_{I} & M_{R}
\end{array}\right]\succcurlyeq0,\label{eq:mapping_operator}
\end{equation}
where $M_{R}$ and $M_{I}$ are the real and imaginary part of the
Hermitian matrix $M=M_{R}+iM_{I}.$ The representation in \ref{eq:mapping_operator}
works even if the matrix $M$ is not PSD. For a quantum state $\vert\psi\rangle=\vert\psi_{R}\rangle+i\vert\psi_{I}\rangle$,
with real part $\vert\psi_{R}\rangle$ and imaginary part $\vert\psi_{I}\rangle,$
one can use the following representation,
\begin{equation}
\vert\psi\rangle\leftrightarrow\left[\begin{array}{cc}
\vert\psi_{R}\rangle & -\vert\psi_{I}\rangle\\
\vert\psi_{I}\rangle & \vert\psi_{R}\rangle
\end{array}\right].\label{eq:mapping_ket}
\end{equation}
Using the aforementioned representations for quantum state and Hermitian
operator, we obtain the following expression for the expectation values,
\begin{equation}
\langle\psi\vert M\vert\psi\rangle=\left[\begin{array}{cc}
\langle\psi_{R}\vert & \langle\psi_{I}\vert\end{array}\right]\left[\begin{array}{cc}
M_{R} & -M_{I}\\
M_{I} & M_{R}
\end{array}\right]\left[\begin{array}{c}
\vert\psi_{R}\rangle\\
\vert\psi_{I}\rangle
\end{array}\right].\label{eq:expectation_real}
\end{equation}
Since the matrices $D$ and $E$ in \ref{eq:P1} are Hermitian, the
above realification works.
\section{Local and Global Optimality\label{sec:Local-and-Global}}
\begin{thm}
Given a tuple $\left(\boldsymbol{\alpha},\lambda\right)$ that is
local minimum and satisfies $H\ge0$ yields the global minimum of
the program P1 in \ref{eq:P1}.
\end{thm}
\begin{proof}
Suppose there exists a tuple $\left(\tilde{\boldsymbol{\alpha}},\tilde{\lambda}\right)$which
satisfies the first and second order conditions and results in $H\geq0$,
and achieves a lower minimum as compared to $\left(\boldsymbol{\alpha},\lambda\right)$
i.e, 
\begin{equation}
\boldsymbol{\alpha}^{T}D\boldsymbol{\alpha}>\tilde{\boldsymbol{\alpha}}^{T}D\tilde{\boldsymbol{\alpha}.}\label{eq:comp_cost_1-1}
\end{equation}
Since $\left(\tilde{\boldsymbol{\alpha}},\tilde{\lambda}\right)$
satisfies the first order conditions and $\left(\alpha,\lambda\right)$
results in $H\geq0$, we have the following,
\[
\tilde{\alpha}^{T}\left(D+\lambda E\right)\tilde{\alpha}\geq0,
\]
\[
\implies\tilde{\alpha}^{T}D\tilde{\alpha}\geq-\lambda\tilde{\alpha}E\alpha
\]
\[
\implies\tilde{\alpha}^{T}D\tilde{\alpha}\geq-\lambda
\]
\begin{equation}
\implies\tilde{\alpha}^{T}D\tilde{\alpha}\geq\alpha^{T}D\alpha.\label{eq:contradiction_1-1}
\end{equation}
The expression in \ref{eq:contradiction_1-1} violates our assumption
i.e., expression in \ref{eq:comp_cost_1-1}. To complete the proof,
we need to show that for the tuples $\left(\alpha_{1},\lambda_{1}\right)$
and $\left(\alpha_{2},\lambda_{2}\right)$ that satisfy the first
and second order conditions and result in $H\geq0,$the associated
objective values are same. Since $D+\lambda_{1}E\geq0$, we have
\[
\alpha_{2}\left(D+\lambda_{1}E\right)\alpha_{2}\geq0
\]
\[
\implies\alpha_{2}D\alpha_{2}\geq-\lambda_{1}
\]
\begin{equation}
\implies\alpha_{2}D\alpha_{2}\geq\alpha_{1}D\alpha_{1}.\label{eq:comparison_1-1}
\end{equation}
Similarly $D+\lambda_{2}E\geq0$ implies
\begin{equation}
\alpha_{1}D\alpha_{1}\geq\alpha_{2}D\alpha_{2}.\label{eq:comparsison_2-1}
\end{equation}
From \ref{eq:comparison_1-1} and \ref{eq:comparsison_2-1}, we have
\[
\alpha_{1}D\alpha_{1}=\alpha_{2}D\alpha_{2}.
\]
\end{proof}
\end{document}